\documentclass{elsart}
\begin{document}
\begin{flushright}
KUPT 2000-01 \\
hep-ph/0001266\\
January 2000 
\end{flushright}


\begin{frontmatter}
\title{$\overline{B}\rightarrow D^{(*)} \pi^-$ beyond naive
factorization in the heavy quark limit}
\author{Junegone Chay}
\address{Department of Physics, Korea University, Seoul 136-701,
Korea}
\begin{abstract}
Nonleptonic decays $\overline{B} \rightarrow D \pi^-$ or
$\overline{B} \rightarrow D^* \pi^-$ are dominated by factorizable
contributions. In the heavy quark limit, nonfactorizable contribution
arises from strong radiative correction and power corrections in
$1/m_b$. I calculate the decay rates for $\overline{B} \rightarrow
D^{(*)} \pi^-$ at next-to-leading order in strong interaction,
including nonfactorizable corrections. The result is expressed in
terms of a convolution of the hard scattering amplitude and the pion
wave function. The decay amplitudes in this method are
independent of the gauge, the renormalization scale, and the
renormalization scheme. The effects of the nonfactorizable
contribution are discussed and numerical estimates are presented. 
\end{abstract}
\end{frontmatter}

It is difficult to understand nonleptonic exclusive decays from first
principles since we do not know nonperturbative effects such as the
soft gluon exchange responsible for the quark confinement. As a first
attempt to understand nonleptonic decays, the idea of
naive factorization was introduced to evaluate the matrix elements of
four-quark operators between hadronic states \cite{bsw}. But it
neither explains experimental data satisfactorily nor is justified
theoretically. It is necessary to include somehow nonfactorizable 
contributions, which arise from gluon
exchange between final-state mesons. There has been a theoretical
improvement on the factorization approximation using the effective
Hamiltonian approach \cite{buras}, and the formalism was improved to
next-to-leading order (NLO) in QCD by phenomenologically introducing
the effective number of colors $N_{\mathrm{eff}}$ \cite{ali}.

It is worth mentioning a technical point in applying the effective
Hamiltonian to calculate decay rates. If we compute the finite part
at NLO with external off-shell quarks \cite{buras}, the decay
amplitudes depend on the gauge choice and the renormalization scheme
\cite{buras2}, hence unphysical. Cheng et al. \cite{cheng} show 
that these dependences are absent if we use on-shell external
quarks. These dependences are artifacts in regulating the infrared
divergence with the off-shell momenta of the external quarks, and they
disappear when we choose on-shell external quarks.

Recently Beneke et al. \cite{beneke} followed the idea used in
exclusive processes by Brodsky and Lepage \cite{brodsky} to calculate
the nonfactorizable contribution in the heavy quark limit. I will employ
the same idea to give an improved analysis on $\overline{B}
\rightarrow D^{(*)} \pi^-$ including nonfactorizable contributions. 

The basic idea is that simplifications occur when the bottom quark
mass $m_b$ and the charm quark mass $m_c$ are large compared to the
strong interaction scale $\Lambda_{\mathrm{QCD}}$ with $r=m_c/m_b$
fixed. In this limit the hadronic matrix elements for $\overline{B}
\rightarrow D^{(*)} \pi^-$ can be written in the form
\begin{eqnarray}
\langle D^{(*)} \pi^-| J_1^{\mu} J_{2\mu} |\overline{B} \rangle &=&
\langle D^{(*)} |J_1^{\mu} |\overline{B}\rangle \langle \pi^-|
J_{2\mu}|0\rangle \nonumber \\
&\times&\Bigl[1+\sum_n A_n \alpha_s^n +
O(\frac{\Lambda_{\mathrm{QCD}}}{m_b},
\frac{\Lambda_{\mathrm{QCD}}}{m_c}) \Bigr],  
\end{eqnarray}
where $J_1$, $J_2$ are the bilinear quark current operators in the
effective weak Hamiltonian. The naive factorization corresponds to
neglecting the $\alpha_s$ corrections and the power corrections in
$\Lambda_{\mathrm{QCD}}$. In this case the matrix element factorizes
into a form factor and a decay constant. If we include higher-order
radiative corrections, this naive factorization is broken. However,
the corrections are calculable from first principles using
perturbation theory. Furthermore the matrix elements of
the current operator between $\overline{B}$ and $D^{(*)}$ states are
also calculable systematically in powers of the inverse heavy quark
masses using the heavy quark effective theory (HQET) \cite{luke}.

The effective weak Hamiltonian for nonleptonic decays such as
$\overline{B}\rightarrow D \pi^-$ or $\overline{B}\rightarrow D^*
\pi^-$ is written as  
\begin{equation}
H_{\mathrm{eff}} = \frac{G_F}{\sqrt{2}} V_{cb} V_{ud}^* (C_1 O_1 +C_2
O_2), 
\label{effh}
\end{equation}
where the four-quark operators $O_1$ and $O_2$ are given by
\begin{eqnarray}
O_1 &=& \Bigl( \overline{c}_{\alpha} \gamma_{\mu} (1-\gamma_5)
b_{\alpha} \Bigr) \Bigl(\overline{d}_{\beta} \gamma^{\mu} (1-\gamma_5)
u_{\beta} \Bigr), \nonumber \\
O_2 &=& \Bigl( \overline{c}_{\beta} \gamma_{\mu} (1-\gamma_5)
b_{\alpha} \Bigr) \Bigl(\overline{d}_{\alpha} \gamma^{\mu} (1-\gamma_5)
u_{\beta} \Bigr).
\end{eqnarray}
In Eq.~(\ref{effh}), $V_{cb}$ and $V_{ud}$ are the
Cabibbo-Kobayashi-Maskawa (CKM) matrix elements, $G_F$ is the Fermi
constant, and $\alpha$, $\beta$ are color indices. The coefficient
functions $C_1$ and $C_2$ are known at NLO.

In the pioneering work of Politzer and Wise \cite{polwi}, they
calculated the ratio of the decay rates $\Gamma (\overline{B}
\rightarrow D \pi^-)/\Gamma  (\overline{B} \rightarrow D^* \pi^-)$
including the perturbative calculation of the nonfactorizable
contribution. Since the coefficient 
functions $C_1$ and $C_2$ were known at leading order at the time,
they could present only the ratio of the decay rates. But now that 
the Wilson coefficients are known at NLO \cite{misiak},
we can calculate the decay rates themselves. Here I present the
NLO result for the decay rates $\Gamma (\overline{B}
\rightarrow D \pi^-)$ and $\Gamma (\overline{B} \rightarrow D^*
\pi^-)$. 

As in other exclusive processes considered by Brodsky and Lepage
\cite{brodsky}, the matrix elements of the operators can be written as
the sum of products of matrix elements between $\overline{B}$ and
$D^{(*)}$ states with the integral over the momentum fraction $x$ of a
hard scattering amplitude times the pion wave function
$\phi_{\pi} (x,\mu)$, normalized as
\begin{equation}
\int_0^1 dx \phi_{\pi} (x,\mu) =1.
\end{equation}
Here $x$ is the momentum fraction of a quark (or an antiquark) in the 
pion with respect to the pion momentum $p_{\pi}^{\mu}$.   
The hard scattering amplitude can be calculated in perturbation
theory. There are other contributions, for example, from the gluon
exchange with a spectator quark. They may not be negligible in
$\overline{B} \rightarrow \pi \pi$ decays, but they are power
suppressed and negligible in this case since the spectator quark is
absorbed by the charm quark.

The Feynman diagrams, which contribute to the nonfactorization, are
shown in Fig.~1. Other Feynman diagrams, in which a gluon is
exchanged between quarks, are attributed to either a meson wave
function or a form factor. Specifically, the contribution of the gluon 
exchange between the quark and antiquark pair forming a pion is
absorbed in the pion wave function, and the contribution of the gluon
exchange between the $b$ and the $c$ quark is absorbed in the form
factor for $\overline{B} \rightarrow D^{(*)}$. 

Since we assign a certain set of Feynman diagrams to nonfactorizable
contributions depending on final-state mesons, it is 
important to specify first which quark and antiquark form a
meson. Only after the specification, we arrange the operators in such
a way that they give correct results. For example, in $\overline{B}
\rightarrow D\pi^-$, we use the operators in Eq.~(\ref{effh}) as they
are. This is called ``charge-retention configuration''. On the other
hand, in $\overline{B} \rightarrow D\pi^0$, we have to rearrange the
operators by Fierz transformation first, and calculate the
corresponding Feynman diagrams like Fig.~1. This is called
``charge-changing configuration''. Therefore the calculation in this
method is process dependent. This prescription sounds trivial in
$\overline{B} \rightarrow D\pi$, but it has a drastic effect on the
processes involving penguin operators. 

The amplitude for $\overline{B}\rightarrow D \pi^-$ can be written in a
compact form as
\begin{eqnarray}
\langle D \pi^- |H_{\mathrm{eff}} | \overline{B} \rangle &=&
\frac{G_F}{\sqrt{2}} V_{cb} V_{ud}^* if_{\pi} p_{\pi}^{\mu} \nonumber
\\
&&\times \Bigl(a_- \langle D
|\overline{c}\gamma_{\mu} (1-\gamma_5) b|\overline{B}\rangle
+ a_+ \langle D|\overline{c}\gamma_{\mu} (1+\gamma_5)
b|\overline{B}\rangle \Bigr),  
\label{ampl}
\end{eqnarray}
where $f_{\pi}$ is the pion decay constant. In Eq.~(\ref{ampl}), the
amplitude is decomposed into two parts in which $a_-$ remains finite,
while $a_+$ approaches zero in the limit $r\rightarrow 0$. 

The form factors for $\overline{B}\rightarrow D^{(*)}$ become
simplified in the limit where $m_b$ and $m_c$ are large compared to
$\Lambda_{\mathrm{QCD}}$, with the ratio $r=m_c/m_b$ fixed. To leading
order in $m_b$ (and $m_c$), the heavy-heavy form factors can be
written as   
\begin{eqnarray}
p_{\pi}^{\mu} \langle D | \overline{c} \gamma_{\mu} (1\pm \gamma_5 )
b|\overline{B} \rangle &=& m_b \Bigl[ (1-r) \langle H_c (v^{\prime}) |
\overline{h}_{v^{\prime}}^{(c)} h_v^{(b)} |H_b (v) \rangle
\nonumber \\
&\mp& (1+r) \langle H_c (v^{\prime}) |
\overline{h}_{v^{\prime}}^{(c)}\gamma_5 h_v^{(b)} |H_b (v)
\rangle \Bigr],
\end{eqnarray}
where $h_{v^{\prime}}^{(c)}$ and $h_v^{(b)}$ are $c$ and $b$ quark
fields in the HQET. We can write a similar form factor for
$\overline{B}\rightarrow D^{*}$. The corrections in powers of
$\Lambda_{\mathrm{QCD}}/m_b$, $\Lambda_{\mathrm{QCD}}/m_c$ can be 
systematically calculated in the HQET \cite{luke}. Also there is a
detailed analysis on the form factor in $\overline{B}\rightarrow
D^{(*)}$ in terms of the Isgur-Wise function at small momentum
transfer $q^2 = m_{\pi}^2$  from the experimental results of
$\overline{B} \rightarrow D\pi$ \cite{mart}.

The coefficients $a_{\pm}$ arise from the QCD corrections and they are
given as, at NLO,
\begin{equation}
a_- = C_1 +\frac{C_2}{N} +\frac{\alpha_s}{4\pi} C_2 \frac{C_F}{N} F,
\ \ a_+ =\frac{\alpha_s}{4\pi}  C_2 \frac{C_F}{N}  G.
\label{apm}
\end{equation}
Here $C_F =(N^2 -1)/(2N)$, and $N=3$ is the number of colors. The
quantities $F$ and $G$ are obtained from Fig.~1 by symmetrizing on
interchange of $x$ with $1-x$. They are written as
\begin{eqnarray}
F &=& -18 +12 \ln \frac{m_b}{\mu} -3i\pi +6\ln (1-r^2) + \int_0^1 dx
\phi_{\pi} (x,\mu) f_1 (x),
\nonumber \\
G &=& -r\int_0^1 dx \phi_{\pi} (x,\mu) g_1 (x).
\end{eqnarray}

The $x$-dependent parts $f_1 (x)$ and $g_1(x)$ have the form
\begin{eqnarray}
f_1 (x)&=& 3\ln x(1-x) \nonumber \\
&&-\frac{3}{1-x(1-r^2)} \Bigl[ \ln x (1-r^2) +
r^2 \ln (1-x)(1-r^2) +i\pi r^2 \Bigr] \nonumber \\
g_1 (x) &=& \frac{1}{1-x(1-r^2)} \Bigl[ 2-\ln x(1-x) -2\ln (1-r^2) +
\ln r^2 -i\pi \Bigr] \nonumber \\
&& +\frac{1}{\Bigl(1-x(1-r^2) \Bigr)^2} \Bigl[ \ln x (1-r^2) + r^2 \ln  
(1-x)(1-r^2) \nonumber \\
&&-r^2 \ln r^2 +i\pi r^2 \Bigr].
\label{fg1}
\end{eqnarray}
Eq.~(\ref{fg1}) is obtained by using the symmetry of the pion wave
function $\phi_{\pi} (x)$ under $x \leftrightarrow 1-x$.  
The hard scattering amplitudes are infrared finite because the
infrared divergence is cancelled when we add all the Feynman diagrams
in Fig.~1. Note that $G\rightarrow0$ as $r$ approaches zero since
$g_1$ behaves as $\ln r$, hence $a_+
\rightarrow 0$. And $F$ becomes identical to the result for 
$\overline{B} \rightarrow \pi \pi$ for $O_1$ and $O_2$ \cite{beneke}.

In obtaining Eq.~(\ref{apm}), I put all the external quarks on the
mass shell and $m_u = m_d =0$. After a straightforward calculation, I
find that there is no gauge dependence with the on-shell external
quarks. At NLO, 
the renormalization scale dependence of the hard-scattering amplitudes
is cancelled by that in the Wilson coefficient
functions. Furthermore, we have the same
result when we use either the NDR scheme or the HV scheme. Therefore
we have the decay amplitudes independent of the choice of the gauge,
the renormalization scale, and the renormalization scheme. The
detailed proof will appear in a forthcoming paper.

It is interesting to note that imaginary parts appear through 
final-state interactions. Specifically the imaginary part comes from
the gluon exchange of a quark and antiquark pair in $\pi$ with the $c$
quark. In $\overline{B} \rightarrow D^{(*)} \pi^-$, the relevant CKM
matrix elements $V_{ud}$ and $V_{cb}$ are almost real. But in other
$B$ decays in which CP violation is studied, this strong phase should
be disentangled before we try to get any information on the CKM matrix
elements. Here I stress that the strong phase is calculable in
perturbation theory. However, there is a caveat in numerical analysis
on the imaginary part since it is sensitive to the renormalization
scale at NLO. The real part is independent of
the renormalization scale since the renormalization scale dependence
of the hard scattering amplitude is cancelled by the Wilson
coefficients at NLO. However, there is no such cancellation for the
imaginary part since the Wilson coefficients at NLO are
real. Therefore the imaginary part is sensitive to the renormalization
scale at this order.

If we use the leading-order pion wave function $\phi_{\pi} (x,m_b) =
6x (1-x)$, we get
\begin{equation}
a_- = 1.062+0.030i, \ \ a_+ = 0.00056 -0.0041i,
\end{equation}
with $\mu = m_b=4.8$ GeV, and $r=0.3$. In the limit $r\rightarrow 0$,
$a_- = 1.064+0.025i$ while $a_+=0$. The coefficients
$a_{\mp} (D\pi)$ are listed for different values of the
renormalization scales in Table~\ref{table1}. The real part is
insensitive to the change of the renormalization scale, while the
imaginary part is not. When $r$ is varied from 0.2 to 0.4 with
$\mu=m_b$ fixed, $a_-$ is in the range $(1.063+0.027i,
1.062+0.033i)$ and $a_+$ is in the range $(-0.00063-0.0035i,
0.0018-0.0041i)$. From this observation, we can conclude that
numerically $a_+$ is negligible compared to $a_-$, and $a_-$ is
insensitive to $r$ for $0.2\leq r\leq 0.4$. $a_-$ increases about 3\%
compared to the leading-order value, verifying that the factorizable
contribution dominates in $\overline{B} \rightarrow D^{(*)} \pi^-$. 

\begin{table}
\caption{The QCD coefficients $a_{\mp} (D\pi^-)$ at NLO with $r=0.3$
for three different renormalization scales $\mu$ with $m_b = 4.8$
GeV. The values in the brackets are those with $r=0$ for
comparison. And the values in the parentheses are the leading order
values.} 

\begin{center}
\begin{tabular}{cccc}\hline 
&$\mu=m_b/2$& $\mu=m_b$& $\mu=2m_b$ \\ \hline
$a_- (D\pi^-)$ & $1.082+0.056i$ & $1.062+0.030i$& $1.043+0.016i$ \\ 
&[$1.084+0.047i$]& [$1.064+0.025i$]&[$1.044+0.014i$] \\
&(1.062)&(1.031)& (1.014) \\
$a_+ (D\pi^-)$ & $0.0010-0.0077i$ & $0.00056-0.0041i$&
$0.00030-0.0022i$ \\
&(0)& (0) &(0) \\  \hline 
\end{tabular}
\end{center}
\label{table1}
\end{table}

In predicting decay rates, the main uncertainty comes from the form
factor for $\overline{B}\rightarrow D^{(*)}$. There are several ways
to parameterize the Isgur-Wise functions and I employ the NRSX model
\cite{nrsx} and the results from Ref.~\cite{mart} to evaluate the
decay rates. The branching ratios are given as
\begin{eqnarray}
\mathrm{Br} (\overline{B} \rightarrow D\pi^-) &=& (3.1 \pm 0.7)
\times 10^{-3}, \nonumber \\
\mathrm{Br} (\overline{B} \rightarrow D^* \pi^-) &=& (2.9 \pm 0.5)
\times 10^{-3},
\end{eqnarray}
which are consistent with the experimental results $(3.1 \pm 0.4 \pm
0.2)\times 10^{-3}$ and $(2.8\pm 0.4 \pm 0.1) \times 10^{-3}$
respectively.

The decay rates $\Gamma (\overline{B} \rightarrow D^{(*)} \pi^-)$ are
parameterized in terms of a single parameter  
$a_1^{\mathrm{eff}}$ in Ref.~\cite{neubert}, which is fit to
experimental data. In this approach, this is not a parameter, but a
process-dependent quantity calculable in perturbation
theory. $|a_1^{\mathrm{eff}}|^2$ 
corresponds to $|a_- +a_+|^2=1.131$ for $\overline{B} \rightarrow
D\pi^-$, and to $|a_- -a_+|^2 =1.128$ for $\overline{B} \rightarrow
D^*\pi^-$. And we do not have to introduce an effective number of
colors $N_{\mathrm{eff}}$ \cite{ali} to parameterize the
nonfactorizable contributions. We fix the number of colors at $N=3$.

Here I present the nonleptonic decay rates for $\overline{B}
\rightarrow D^{(*)} \pi^-$ in the heavy quark limit in which the
nonfactorizable contributions are calculated in perturbation theory at 
NLO. This calculation is an improved one compared to previous
results since we include the momentum-dependent part in the
decay amplitudes, which was not included so far,  as well as the
momentum-independent part. However, the result still depends on
models. The decay rates depend on the pion wave functions
which reflect our ignorance of purely nonperturbative effects. If we
consider pions only, there are other independent studies on the shape
of the pion wave functions using the QCD sum rules, and the light-cone
formalism, etc. \cite{chernyak,brod2}. Therefore we can be 
reasonably sure of the form of the pion wave function. But if the
light mesons in the final state include $K$, $K^*$, $\eta$,
$\eta^{\prime}$, or $\rho$ in this
framework, the model dependence on the meson wave functions can
be severe. In this case, the experimental data should be used as an
input to determine the wave functions and predict other decay modes.

The main point of this paper is to show that nonfactorizable
contributions are calculable from first principles using perturbation 
theory in the heavy quark limit. However, in color-allowed decays such
as $\overline{B}
\rightarrow D^{(*)} \pi^-$, the nonfactorizable contribution is
small. In color-suppressed decays such as $\overline{B} \rightarrow
D\pi^0$, it will be significant. A complete analysis for
$\overline{B}\rightarrow D\pi$ will be presented in a future
publication.   
It will be also interesting to apply this technique to various
$B$ decays such as the decays into two light mesons. An extensive
study on the $B$ decays into two light mesons is in progress. 

\section*{Acknowledgements}
The author was supported in part by the Ministry of Education grants
KRF-99-042-D00034 D2002, BSRI 98-2408, and the German-Korean
scientific exchange program DFG-446-KOR-113/72/0. The author would
like to thank Ahmed Ali, Christoph Greub, and Pyungwon Ko for
stimulating discussions.

\newpage

Figure caption

Figure 1: Feynman diagrams for nonfactorizable contribution. The dots
denote the operator $O_2$.

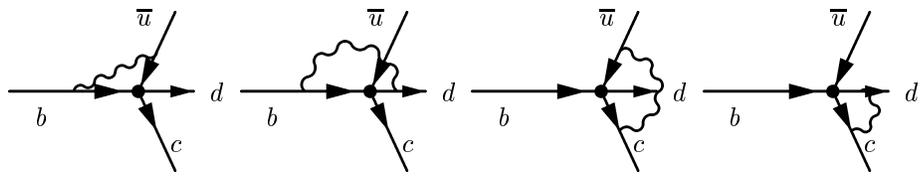
\end{document}